\documentclass[acmsmall,nonacm]{acmart}
\AtBeginDocument{%
  }

\setcopyright{acmlicensed}
\copyrightyear{2025}
\acmYear{2025}
\acmDOI{XXXXXXX.XXXXXXX}
\acmConference[Conference acronym 'XX]{Make sure to enter the correct
  conference title from your rights confirmation email}{June 03--05,
  2018}{Woodstock, NY}
\acmISBN{978-1-4503-XXXX-X/2018/06}

\usepackage{multirow}
\usepackage{adjustbox}
\usepackage{booktabs}
\usepackage{subfig}
\usepackage[most]{tcolorbox}
\usepackage[table]{xcolor}
\usepackage{enumitem}
\definecolor{myStart}{RGB}{255,180,180}  % 浅粉色

\newcommand{\ceco}[1]{%
    \pgfmathsetmacro{\val}{#1}%
    \pgfmathsetmacro{\tmp}{10 + ((\val - 34.8)/(100-34.8))*80}%
    \pgfmathtruncatemacro{\colorpercent}{\tmp}%
    \edef\temp{\noexpand\cellcolor{myStart!\colorpercent}{#1}}%
    \temp
}
\begin{document}
% \title{\texttt{ReMind}: Understanding and Mitigating the Limitations of LLMs in Deductive Code Reasoning}
\title{\texttt{ReMind}: Understanding Deductive Code Reasoning in LLMs}

% \affiliation{%

\author{Jun Gao}
\affiliation{%
  \institution{Zhejiang University}
  \city{Hangzhou}
  \country{China}}
\email{jgao1106@zju.edu.cn}

\author{Yun Peng}
\affiliation{%
  \institution{The Chinese University of Hong Kong}
  \city{HongKong}
  \country{China}}
\email{jgao1106@zju.edu.cn}

\author{Xiaoxue Ren}
\affiliation{%
  \institution{Zhejiang University}
  \city{Hangzhou}
  \country{China}}
\email{xxren@zju.edu.cn}

% \affiliation{%

% \affiliation{%

% \affiliation{%

% \affiliation{%

% \affiliation{%

% \affiliation{%

\renewcommand{\shortauthors}{Paper Under Review}

\begin{abstract}
Large Language Models (LLMs) have achieved remarkable progress in code-related tasks. 
Despite their advancement, empirical evidence reveals that they still struggle with \emph{deductive code reasoning}, the ability to reason about the program execution process.
While prior studies have recognized this limitation, the underlying causes remain largely underexplored.
In this paper, we begin by presenting a comprehensive empirical study that reveals three key challenges undermining deductive code reasoning: (1) an intrinsic gap between generation and reasoning abilities, (2) a consistent bias towards code sources, and (3) weak zero-shot generalization on complex benchmarks.
In light of these challenges, we propose \texttt{ReMind}, a multi-agent framework composed of \texttt{Mutator}, \texttt{Executor}, and \texttt{Inspector}.
The \texttt{Mutator} generates code variants to mitigate bias towards code sources, the \texttt{Executor} traces variable states step-by-step to expose inconsistency, and the \texttt{Inspector} identifies problematic reasoning steps and provides control-flow refinement to bridge the intrinsic reasoning gap. 
Through their coordinated collaboration, \texttt{ReMind} systematically identifies and refines reasoning flaws, achieving outstanding performance and enabling robust zero-shot generalization.
Extensive experiments on two benchmarks with five LLMs demonstrate the superior advantages of \texttt{ReMind} compared to baseline approaches in deductive code reasoning.
\end{abstract}

% <concept_id>10011007.10010940</concept_id>
% <concept_desc>Software and its engineering~Software organization and properties</concept_desc>

% \ccsdesc[500]{Software and its engineering~Software organization and properties}


\maketitle

\section{Introduction}
Large Language Models (LLMs) have achieved impressive progress in code-related tasks such as code generation~\cite{chen2021evaluating, hurst2024gpt, liu2024deepseek, guo2025deepseek}, completion~\cite{sun2025don}, and repair~\cite{yang2024swe, bouzenia2024repairagent, jin2023inferfix}.
Despite their apparent success that LLMs have increasingly been integral to software development workflows, empirical evidence indicates that even state-of-the-art (SoTA) LLMs often struggle with \emph{deductive code reasoning}~\cite{haroon2025accurately, lichain, zhaounveiling}.
The ability of LLMs to reliably play the role of an interpreter or compiler, tracking dynamic variable states step-by-step before arriving at the final correct result, without any interaction with actual execution environments.
Such deficiencies pose a critical barrier to applications that require trustworthy code reasoning, such as automated debugging~\cite{malik2009case, tian2024debugbench}, program verification~\cite{wen2024enchanting, bhatia2024verified}, and AI-assisted software development~\cite{maninger2024towards}.

Unfortunately, existing studies on deductive code reasoning~\cite{lichain,zhaounveiling} rarely investigate the underlying causes of reasoning failures in LLMs.
This leaves open the fundamental question: \emph{What factors limit LLMs to reliably perform deductive code reasoning?}
Rather than attributing failures solely to model capacity, we empirically examine three practical factors that may undermine the reliability of deductive code reasoning: (1) the gap between generation and reasoning abilities, (2) reasoning biases of LLMs shaped by pretraining corpora, and (3) the impacts of complex benchmarks in zero-shot settings.
To systematically study these factors, we conduct an empirical study guided by the following reasoning questions:
\begin{itemize}
    \item \textbf{RQ1:} To what extent can LLMs accurately reason about the execution path of code they generate? 
    \item \textbf{RQ2:} How consistent are LLMs when reasoning over code from diverse sources (e.g., generated by different LLMs)?
    \item \textbf{RQ3:} How effectively can LLMs generalize to complex and realistic benchmarks (e.g., LiveCodeBench~\cite{jainlivecodebench}) in zero-shot settings? 
\end{itemize}
However, answering these research questions is non-trivial.
Standard code reasoning benchmarks, such as CruxEval~\cite{gu2024cruxeval}, predominantly feature short, simple code snippets, limited test cases, and fixed code sources.
These characteristics fall short for our empirical study and rarely capture real-world scenarios.
To isolate the deductive reasoning ability of LLMs from implementation and code functionality, we adopt a rigorous validation protocol on two code generation benchmarks (i.e., HumanEval~\cite{chen2021evaluating} and LiveCodeBench~\cite{jainlivecodebench}), which offer more comprehensive test cases.
This protocol first prompts all LLMs to generate candidate code solutions for a given code generation problem, then validates them against official test cases.
The protocol retains only those cases where \emph{all} LLMs produce functionally correct code, i.e., pass every test case in actual execution environments.
As a result, the protocol produces a well-designed evaluation benchmark tailored to the three research questions, where all codes are generated by LLMs under study (\emph{RQ1}) and are functionally equivalent yet source-diverse (\emph{RQ2}).
Moreover, the benchmark encompasses both basic questions from HumanEval and challenging ones from LiveCodeBench (\emph{RQ3}).
Our empirical evaluation across five LLMs on two benchmarks reveals three core findings:
\begin{itemize}
    \item \textbf{Intrinsic Reasoning Gap}: Even when LLMs generate functionally correct code, they frequently fail to accurately deduce the output of code they generate, suggesting a fundamental gap (up to 44.7\%) between generation and deductive reasoning abilities.
    
    \item \textbf{Self-Reasoning Bias:} LLMs exhibit a strong performance drop (losing up to 43\% relative performance) when reasoning about code from different sources, indicating a significant bias toward their own or familiar coding patterns.

    \item \textbf{Zero-shot Reasoning Dilemma:} In zero-shot settings, deductive reasoning on complex, real-world codes (LiveCodeBench) remains markedly unreliable compared to simpler ones, highlighting limited generalization beyond basic benchmarks (HumanEval).

\end{itemize}

Building on these insights, we accordingly propose \texttt{ReMind}, a novel multi-agent framework for enhancing deductive code reasoning by integrating code mutation, execution, and inspection into a collaborative loop. 
Specifically, \texttt{ReMind} is viewed as a form of \textit{test-time scaling}~\cite{jaech2024openai,muennighoff2025s1,zeng2025revisiting,gao2025uniicl,gao2025aim,gao2025interleaved}, where reasoning reliability is improved through structured computation at inference time without requiring training.
At a high level, \texttt{ReMind} is composed of three agents: \texttt{Mutator}, \texttt{Executor}, and \texttt{Inspector}.
To mitigate \emph{self-reasoning bias}, the \texttt{Mutator} generates code variants with both semantic and structural mutations.
The resulting mutated implementations may deviate functionally from the original code since the Mutator is LLM-driven.
Instead of treating such deviations as intolerable errors, \texttt{ReMind} considers them as meaningful signals.
Such deviations highlight error-prone code segments that are difficult to handle, providing actionable insights for subsequent \texttt{Executor} and \texttt{Inspector}.
The \texttt{Executor} then acts as an interpreter or compiler, reasoning about the execution paths of both the original code and the mutated variants, tracing the variable outputs step-by-step before arriving at the final outputs.
Deviations introduced by the \texttt{Mutator} potentially lead to divergent execution paths and inconsistent final outputs.
Finally, the Inspector validates inconsistent execution trajectories against corresponding control-flow graphs (CFGs) to locate code regions that exhibit deviations.
Then, \texttt{Inspector} identifies erroneous reasoning steps and accordingly provides corrective feedback to the \texttt{Executor}, narrowing the \emph{intrinsic reasoning ability gap}.
By orchestrating these agents in a closed-loop framework, \texttt{ReMind} systematically exposes, identifies, and refines reasoning flaws, establishing outstanding reasoning abilities.

Extensive experiments on two benchmarks with five LLMs demonstrate the superior advantages of \texttt{ReMind} in both reasoning performance and robustness.
\texttt{ReMind} achieves accurate predictions irrespective of code origin and reliable zero-shot reasoning, with a maximum average improvement of 23.2 in absolute accuracy over the best performing baselines.%提升了多少
Additionally, the ablation study validates the design of \texttt{ReMind}, while each component offers individual benefits, and their collaboration further yields performance gains.

In summary, our contributions are threefold:
\begin{itemize}
\item We present the first empirical investigation into the underlying causes of LLMs falling short in deductive code reasoning, emphasizing three key challenges, including \emph{intrinsic reasoning gap, self-execution bias, and the zero-shot reasoning dilemma}.

\item Motivated by challenges probed in the empirical study, we accordingly propose \texttt{ReMind}, a multi-agent framework that integrates code mutation, execution process reasoning, and CFG-driven inspection, demonstrating superior advantages in both performance and robustness.

\item We conduct extensive experiments to evaluate the effectiveness of our \texttt{ReMind} using five popular LLMs on both basic and complex benchmarks. Experiments demonstrate consistent improvements in accuracy, stability, and zero-shot generalization regardless of code origin compared to existing baselines.
\end{itemize}

\section{Empirical Study}
Empirical evidence has indicated that even SoTA LLMs remain fall short in \textit{deductive code reasoning}~\cite{haroon2025accurately,lichain,zhaounveiling}.
While prior studies demonstrate surface-level limitations, the underlying causes remain largely underexplored, raising a critical open question: \emph{What factors limit LLMs to reliably perform deductive code reasoning?}
To shed light on this question, we conduct this empirical study to examine three potential factors that may undermine reasoning reliability: (1) the gap between generation and reasoning abilities, (2) reasoning biases of LLMs shaped by pretraining corpora, and (3) the impacts of complex benchmarks in zero-shot settings.

% \begin{itemize}
%     \item \textbf{RQ1:} To what extent can LLMs accurately reason about the execution path of code they generate? 
%     \item \textbf{RQ2:} How consistent are LLMs when reasoning over code from diverse sources (e.g., generated by different LLMs)?
%     \item \textbf{RQ3:} How effectively can LLMs generalize to \emph{complex and realistic benchmarks} (e.g., LiveCodeBench~\cite{jainlivecodebench}) in zero-shot settings? 
% \end{itemize}

The remainder of this section lays the foundation for our study.
We first formally define the problem of deductive code reasoning and introduce the corresponding empirical setup.
To shed light on the challenges involved, we present quantitative experiments using two straightforward yet illustrative examples that exemplify common code reasoning errors.

\subsection{Problem Definition of Deductive Code Reasoning}
\emph{Deductive Code Reasoning} is a crucial capability in real-world applications such as automated code debugging~\cite{malik2009case,tian2024debugbench}, program verification~\cite{wen2024enchanting,bhatia2024verified}, and AI-assisted software development~\cite{maninger2024towards}.
It refers to the task where, given a program $\mathcal{P}$ and its input $\mathcal{I}$, an LLM executor $E$ attempts to predict the program’s actual execution output $\mathcal{O}$ purely through its reasoning capabilities~\cite{gu2024cruxeval,jainlivecodebench}, without access to an actual execution environment.
In this setting, the LLM functions like a compiler or interpreter~\cite{kim2024llm}, simulating step-by-step to deduce its final output.
Formally, the executor $E$ takes program $\mathcal{P}$ and input $\mathcal{I}$ as input and produces a predicted output $E(\mathcal{P, I}) \longrightarrow \tilde{\mathcal{O}}$, where $\tilde{\mathcal{O}}$ should match the true output $\mathcal{O}$.
For example, given the program: \texttt{values = [num + 1 for num in numbers]} and the input: \texttt{numbers = [1, 2, 2, 1]}, the correct predicted output would be: \texttt{values = [2, 3, 3, 2]}.

\subsection{Empirical Setup}
To investigate the above research questions, we evaluate a spectrum of SoTA LLMs, including \texttt{GPT-4o-mini}~\cite{hurst2024gpt}, \texttt{DeepSeek-V3}~\cite{liu2024deepseek}, \texttt{DeepSeek-R1}~\cite{guo2025deepseek}, and \texttt{o1} with both high and low reasoning effort~\cite{jaech2024openai}. 
In the empirical study, all models are prompted with simple while widely used \emph{Chain-of-Thought Prompting (CoT)}~\cite{wei2022chain} to elicit reasoning processes~\cite{lichain,jainlivecodebench,zhaounveiling,haroon2025accurately}.
We recognize that standard code reasoning benchmarks typically feature short, simple code snippets, limited test cases, and fixed code sources.
As such, they may be inadequate to reveal the subtle and challenging factors in deductive code reasoning.
To better reflect practical scenarios, we instead use two widely adopted code generation benchmarks, which offer more comprehensive test cases:
\begin{itemize}
    \item  \textbf{HumanEval~\cite{chen2021evaluating}} consists of 164 Python programming tasks designed to assess functional correctness. Its problems are relatively short and focus on algorithmic reasoning.
    
    \item  \textbf{LiveCodeBench~\cite{jainlivecodebench}} is a \emph{continuously updated} benchmark that collects competitive programming problems from platforms such as LeetCode, AtCoder, and Codeforces.
    In this study, we adopt release\_v5, which contains 880 problems released between \emph{May 2023 and Jan 2025}. Our analysis focuses on the subset from \emph{Oct 2023 to Jan 2025 (later than the knowledge cutoff of OpenAI o1), thus mitigating potential data leakage.}
    
\end{itemize}

\begin{table}[!t]
  \centering
  \caption{Statistical details of the used public benchmarks in our study. The question indicates the number of different questions and the Avg. Test indicates the average number of test cases used for execution.}
    \begin{tabular}{lcc}
    \toprule
    \textbf{Benchmark} & \textbf{Questions} & \textbf{Avg. Tests} \\
    \midrule
    \textbf{HumanEval}~\cite{chen2021evaluating} & 152  & 6.9  \\
    \textbf{LiveCodeBench}~\cite{jainlivecodebench} & 328  & 14.6  \\
    \bottomrule
    \end{tabular}
  \label{tab:dataset}
\end{table}

To ensure that errors arise from the deductive code reasoning ability of LLMs rather than implementation and functionality, we adopt a rigorous validation protocol.
This protocol enables a focused analysis of the underlying causes tailored to our three research questions.
Protocol first prompts LLMs under study to generate candidate code snippets and then rigorously validates their syntax and functionality against the corresponding test cases.
Only those instances where \emph{all} LLMs produce correct solutions that pass \emph{every} test case in \emph{actual execution environments} are retained.
This results in a carefully designed evaluation benchmark where code samples generated by all LLMs under study are functionally equivalent but originate from different sources.
Additionally, this benchmark includes both basic problems from HumanEval and more challenging zero-shot problems from LiveCodeBench.
After applying this protocol, we obtain filtered subsets of 152 problems for HumanEval and 328 problems for LiveCodeBench, and we cap the number of test cases at 15 to balance coverage and computational cost. 
Tab.~\ref{tab:dataset} summarizes the dataset statistics used in our empirical study.
Given that each instance includes multiple test cases, we adopt a modified \texttt{Pass@1}-like accuracy metric defined as:
\begin{equation}
\mathrm{Accuracy} = \frac{1}{N}\sum_{i=1}^N \prod_{j=1}^{m_i} [O_{ij} = \tilde{O}_{ij}],
\label{eq:metric}
\end{equation}
where $N$ denotes the total number of instances in the benchmark, $m_i$ is the number of test cases for the $i$-th instance, and $[\cdot]$ is the indicator function that evaluates to $1$ if the condition inside holds and $0$ otherwise.

Building upon this benchmark, we design three crafted experimental scenarios to guide our investigation.
The first is \textbf{Self-Execution} for \emph{RQ1}, where LLMs reason about code generated by themselves, quantifying the gap between code generation and deductive code reasoning abilities.
Second, we define \textbf{Cross-Execution} for \emph{RQ2}, in which the LLMs reason about code produced by other LLMs, evaluating the performance stability of reasoning over different code sources.
Finally, we introduce \textbf{Zero-shot Reasoning} where evaluations are conducted on LiveCodeBench, which explicitly mitigates data leakage through knowledge cutoff of LLMs, providing a rigorous assessment of the generalization capability of LLMs in deductive code reasoning.
Taken together, these scenarios form a comprehensive evaluation for analyzing the underlying causes leading to deductive code reasoning failures.

\subsection{Empirical Results}
\label{sec:pre-res}
\begin{table}[!t]
  \centering
\caption{Comparison of reasoning accuracy across five LLMs on the HumanEval. Mutation indicates that the code is mutated by the underlying LLM of the executor before applying CoT reasoning. DS-V3(R1) is the abbreviation of DeepSeek-V3(R1), and o1-Low (High) indicates OpenAI o1 with low (high) reasoning efforts.}

\begin{tabular}{clccccc|c}
\bottomrule
\multirow{1}{*}{\textbf{Executor}} & \multicolumn{1}{c}{\multirow{2}{*}{\textbf{Method}}} & \multicolumn{6}{c}{\textbf{Benchmark: HumanEval}}  \\
\cline{3-8}   \textbf{LLM}    &       & 4o-mini & DS-V3 & DS-R1 & o1-Low & o1-High & Avg.   \\
\hline
\multirow{2}[1]{*}{\textbf{4o-mini}} & CoT   & 55.3  & 48.7  & 51.3  & 52.6  & 53.9  & 52.4  \\
      & Mutation & -     & 53.9  & 47.4  & 53.9  & 59.2  & 53.6   \\
\hline
\multirow{2}[1]{*}{\textbf{DS-V3}} & CoT   & 53.9  & 85.5  & 52.6  & 48.7  & 55.3  & 59.2  \\
      & Mutation & 61.8  & -     & 56.6  & 53.9  & 59.2  & 57.9   \\

\hline
\multirow{2}[0]{*}{\textbf{DS-R1}} & CoT   & 67.1  & 89.5  & 92.1  & 84.2  & 88.2  & 84.2   \\
      & Mutation & 78.9  & 90.8  & -     & 85.5  & 90.8  & 86.5   \\
\hline
\multirow{2}[0]{*}{\textbf{o1-Low}} & CoT   & 76.3  & 81.6  & 84.2  & 96.1  & 92.1  & 86.1   \\
      & Mutation & 80.3  & 81.6  & 88.1  & -     & 96.1  & 86.5   \\
\hline
\multirow{2}[1]{*}{\textbf{o1-High}} & CoT   & 77.6  & 78.9  & 86.8  & 88.2  & 97.4  & 85.8   \\
      & Mutation & 77.6  & 85.5  & 89.5  & 93.4  & -     & 86.5   \\
\toprule
\end{tabular}

\label{tab:empirical_results}
\end{table}
\begin{figure}[htbp]
    \centering
    \subfloat[Execution accuracy on HumanEval]{
        \includegraphics[width=0.56\textwidth]{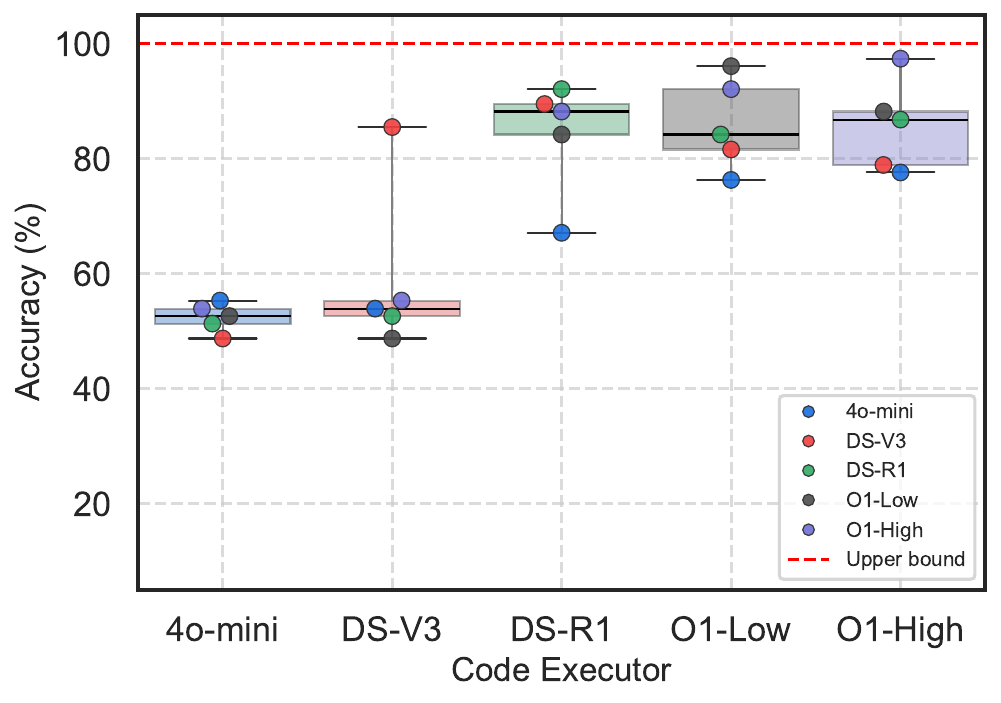}
        \label{fig:box2}
    }
    \hfill
    \subfloat[Relative performance on HumanEval.]{
        \includegraphics[width=0.4\textwidth]{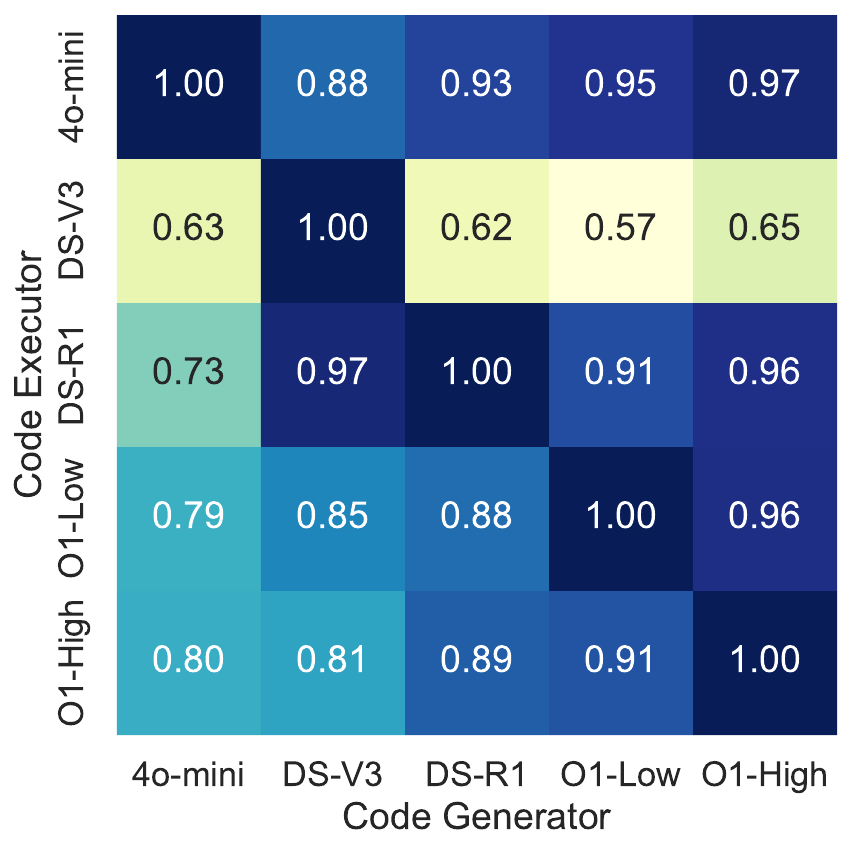}
        \label{fig:box1}
    }
    \caption{Code reasoning accuracy on HumanEval across different LLMs.
    (a) Boxplots show the distribution of execution accuracy across five LLMs, with individual points denoting performance when executing code generated by each corresponding model. 
    The red dashed line indicates the target reasoning performance (upper bound = 100\%).
    (b) Heatmaps illustrate cross-execution performance. Rows correspond to reasoning LLMs and columns to different code sources. Each cell reports the relative accuracy compared to Self-Execution settings (diagonal = 1.00).}
    \label{fig:humaneval-visual}
\end{figure}
\textbf{RQ1: To What Extent Can LLMs Accurately Reason about the Execution Path of Code They Generate?}
While recent LLMs can generate functionally correct code, it remains unclear whether they can reliably reason about the behavior of the code they produce.
Investigating this ability is crucial for understanding the extent to which LLMs possess genuine deductive code reasoning skills compared to code generation abilities.

Tab.~\ref{tab:empirical_results} reports accuracy on HumanEval across five LLMs in the \textbf{Self-Execution} settings, where each LLM is tasked with reasoning about its own generated code using CoT prompting.
Despite all models producing functionally correct code, none achieve perfect reasoning performance, revealing a fundamental gap between their outperforming code generation capabilities and their ability to perform deductive code reasoning.
For example, powerful models such as DeepSeek-V3 and DeepSeek-R1 reach 85.5\% and 92.1\% correctness, respectively.
4o-mini lags far behind at only 55.3\% in HumanEval, indicating a maximum gap of 44.7\% between code generation and deductive code reasoning abilities.

Fig.~\ref{fig:humaneval-visual}(a) complements the tabular results with a visual comparison.
The red dashed line marks the ideal upper bound of 100\% correctness, visualizing the non-negligible gap.
This gap further highlights a critical yet underexplored intrinsic limitation overlooked previously that LLMs struggle to reliably reason about code execution trajectories, even when capable of generating them.
% Additionally, the spread of dots in Fig.~\ref{fig:livecodebench-visual}(a) illustrates notable variance in predictions when models reason about codes generated by other LLMs, a phenomenon we further explore in \emph{RQ2}.

\begin{tcolorbox}[breakable,width=\linewidth-2pt,boxrule=0pt,top=2pt, bottom=2pt, left=2pt,right=2pt, colback=gray!20,colframe=gray!20]
\textbf{Answer to RQ1:}
LLMs demonstrate a gap between their code generation and deductive code reasoning abilities.
The non-negligible gap reveals intrinsic reasoning limitations of LLMs in reliably reasoning about code execution trajectories.
\end{tcolorbox}

\textbf{RQ2: How Consistent Are LLMs When Reasoning over Code from Diverse Sources} 
Previous studies~\cite{jainlivecodebench,zhaounveiling,lichain} have largely overlooked a key challenge in real-world that code snippets can originate from a variety of sources, ranging from human experts to a spectrum of LLMs.
This diversity perhaps increases complexity for reasoning, since LLMs expect to capture the semantics and logic of arbitrary code during reasoning and handle stylistic differences shaped by the training biases of each LLM generator~\cite{li2025preference} simultaneously.
% To investigate this, we formally define \textbf{Cross-Execution} as the scenario where the code generator and the executor are different LLMs.
% This setting enables us to assess how well LLMs generalize their reasoning to functionally equivalent code variants produced by other models.

Tab.~\ref{tab:empirical_results} clearly reveals a consistent \emph{Self-Reasoning Bias} in \textbf{Cross-Execution} settings that every LLM achieves its highest accuracy when reasoning about its own generated code, but suffers noticeable degradation when executing code produced by others.
This effect is also visible in Fig.~\ref{fig:humaneval-visual}(b), where diagonal entries (\textbf{Self-Execution}) are consistently higher than their off-diagonal counterparts.
Among all models, the DeepSeek-V3 executor shows the most dramatic decline in cross-execution settings, losing up to 43\% performance in reasoning about code generated by o1-Low.
While it achieves 85.5\% accuracy in \textbf{Self-Execution}, its performance drops sharply to around 48–55\% when executing code generated by other models. 
Closer inspection suggests that DeepSeek-V3 struggles to robustly handle heterogeneous coding styles established by other LLMs, as exemplified in Motivating Example 2 (Fig.~\ref{fig:motivated}).
Specifically, for some instances, DeepSeek-V3 correctly deduces the outputs for most test cases, but a small subset of failures causes the entire instance to be marked as incorrect under the accuracy metric in Eq.~\ref{eq:metric}.
This sensitivity to partial mismatches prevents DeepSeek-V3 from achieving comparable accuracy.

To further verify this bias effect, we introduce the \emph{Mutation} setting in Tab.~\ref{tab:empirical_results}, where the executor’s underlying LLM explicitly mutates the generated code before reasoning.
This explicitly disrupts style-related factors across different LLMs as reasoned codes are rewritten by the LLM of Executor.
By comparing Mutation with CoT in Tab.~\ref{tab:empirical_results}, we can find that Mutation partially alleviates the self-reasoning bias, resulting in consistent performance improvements in cross-execution settings. 
These findings prove that the observed performance degradation primarily stems from a strong bias toward self-generated code.

\begin{tcolorbox}[breakable,width=\linewidth-2pt,boxrule=0pt,top=2pt, bottom=2pt, left=2pt,right=2pt, colback=gray!20,colframe=gray!20]
\textbf{Answer to RQ2:} LLMs exhibit a consistent self-reasoning bias, where cross-execution performance marginally underperforms compared to self-execution settings.
This performance gap is partially mitigated through mutation, suggesting limited stability of LLMs when reasoning over different code sources.
\end{tcolorbox}

\textbf{RQ3: How Effectively Can LLMs Generalize to Complex and Realistic Benchmarks in Zero-shot Settings?}
\begin{table}[!t]
  \centering
\caption{Comparison of accuracy across different executors on the LiveCodeBench benchmark.}
\begin{tabular}{clccccc|c}
\bottomrule
\multirow{2}{*}{\textbf{Reasoner}} & \multicolumn{1}{c}{\multirow{2}{*}{\textbf{Method}}}  & \multicolumn{6}{c}{\textbf{Benchmark: LiveCodeBench}} \\
\cline{3-8}      &       & 4o-mini & DS-V3 & DS-R1 & o1-Low & o1-High & Avg. \\
\hline
\multirow{2}[1]{*}{\textbf{4o-mini}} & CoT    & 42.1  & 34.8  & 40.2  & 36.6  & 40.8  & 38.9 \\
      & Mutation   & -     & 35.4  & 40.9  & 41.5  & 42.7  & 40.1 \\
\hline
\multirow{2}[1]{*}{\textbf{DS-V3}} & CoT    & 39.0  & 48.8  & 41.5  & 36.9  & 42.7  & 41.8 \\
      & Mutation   & 47.6  & -     & 43.3  & 47.6  & 50.6  & 47.3 \\

\hline
\multirow{2}[0]{*}{\textbf{DS-R1}} & CoT    & 50.0  & 54.3  & 57.3  & 54.3  & 54.9  & 54.1 \\
      & Mutation   & 53.7  & 55.5  & -     & 54.9  & 56.7  & 55.2 \\
\hline
\multirow{2}[0]{*}{\textbf{o1-Low}} & CoT   & 45.7  & 56.7  & 56.7  & 61.6  & 57.9  & 55.7 \\
      & Mutation  & 47.0  & 59.1  & 57.9  & -     & 59.8  & 55.9 \\
\hline
\multirow{2}[1]{*}{\textbf{o1-High}} & CoT    & 54.9  & 57.3  & 59.1  & 60.1  & 64.0  & 59.1 \\
      & Mutation  & 56.1  & 59.8  & 60.4  & 62.8  & -     & 59.8 \\
\toprule
\end{tabular}
\label{tab:empirical_results_lcb}
\end{table}
\begin{figure}[!t]
    \centering
    \subfloat[Execution accuracy on LiveCodeBench.]{
        \includegraphics[width=0.56\textwidth]{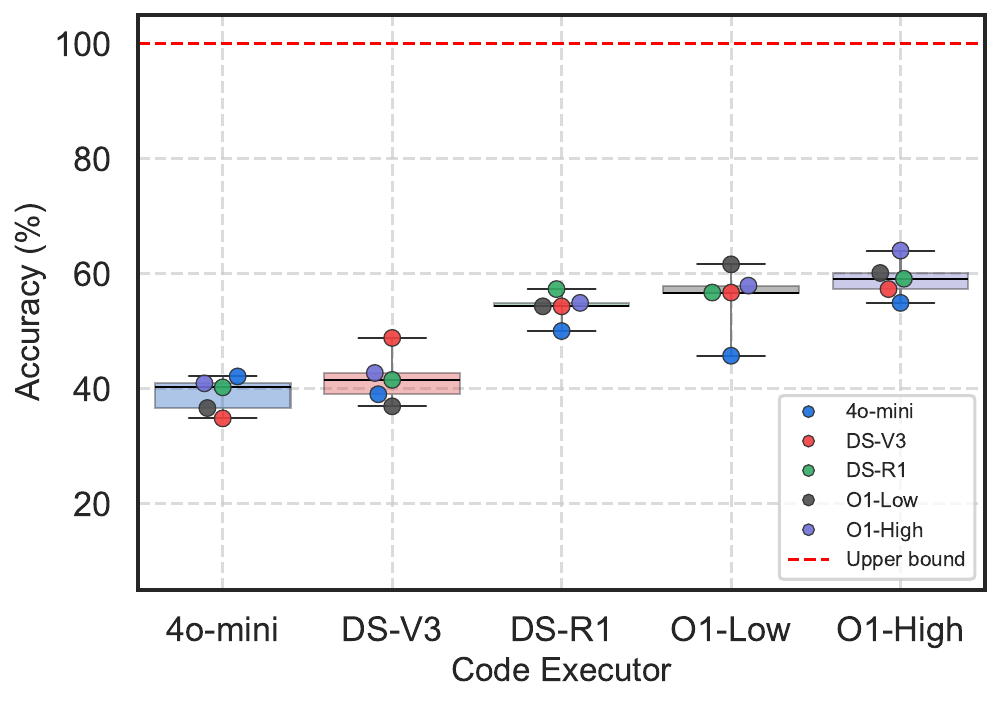}
        \label{fig:heat2}
    }
    \hfill
    \subfloat[Relative performance on LiveCodeBench.]{
        \includegraphics[width=0.4\textwidth]{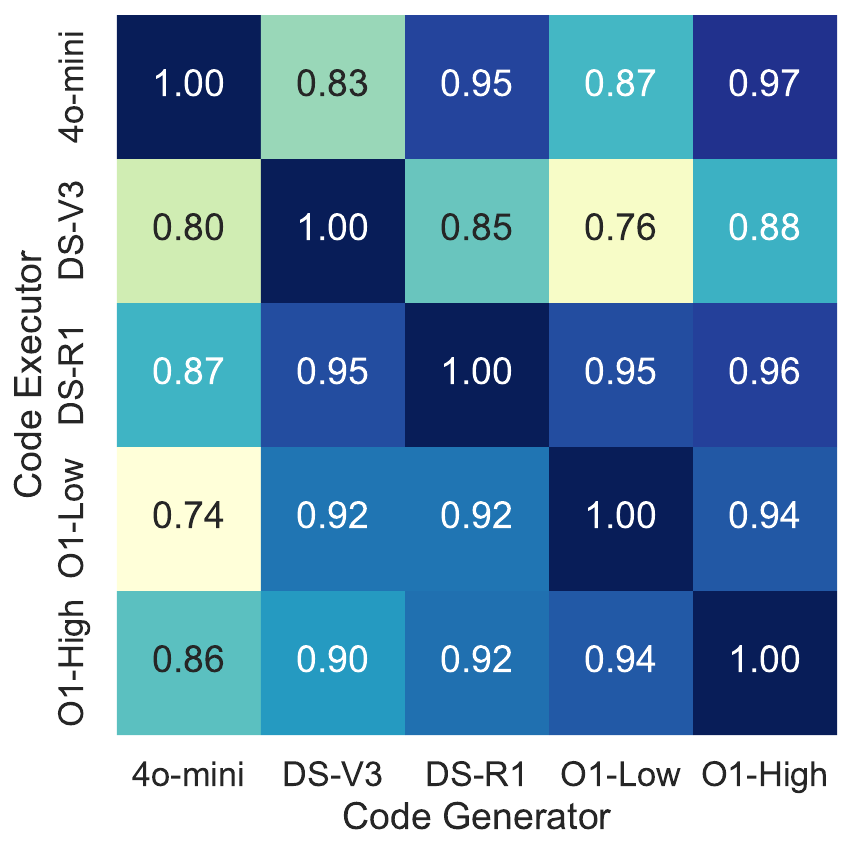}
        \label{fig:heat1}
    }
    \caption{Code execution accuracy boxplots (a) and cross-execution performance heatmaps (b) on LiveCodeBench across different underlying LLMs.}
    \label{fig:livecodebench-visual}
\end{figure}
HumanEval has been publicly available for several years, raising concerns that potential data leakage may compromise the authenticity of reported findings.
To address this concern and strengthen our insights from \emph{RQ1} and \emph{RQ2}, we extend our basic evaluation to \textbf{Zero-shot} reasoning settings.
% LiveCodeBench benchmark provides a compelling zero-shot test of reasoning ability, as the evaluated models are restricted by their knowledge cutoff.

As shown in Tab.~\ref{tab:empirical_results_lcb}, accuracy on LiveCodeBench drops sharply across all executors compared to HumanEval in Tab.~\ref{tab:empirical_results}.
Even the top-performing o1-High model, which exceeds 90\% accuracy on HumanEval self-execution, attains only 64\% under the same setting.
DeepSeek-R1 exhibits a comparable decline, dropping from an average of 84.2\% accuracy on HumanEval to only 54.1\% on LiveCodeBench, while smaller models, such as 4o-mini, fall from an average of 52\% to 34–42\%. 
These results highlight the substantial challenge posed by complex, realistic tasks, even when models have produced them.

Boxplots in Figure~\ref{fig:livecodebench-visual}(a) further illustrate this trend that the overall distribution of execution accuracy is markedly lower than in HumanEval (Figure~\ref{fig:humaneval-visual}(a)), with all models performing well below the ideal upper bound.
Similarly, the heatmap (Figure~\ref{fig:livecodebench-visual}(b)) shows that cross-execution performance on LiveCodeBench is consistently weaker than self-execution, reflecting the same \emph{self-reasoning bias} observed previously.

\begin{tcolorbox}[breakable,width=\linewidth-2pt,boxrule=0pt,top=2pt, bottom=2pt, left=2pt,right=2pt, colback=gray!20,colframe=gray!20]
\textbf{Answer to RQ3:} LLMs face a zero-shot reasoning dilemma, exhibiting substantial performance degradation when evaluated on complex and realistic benchmarks.
This decline reveals limited generalization of LLMs beyond basic benchmarks.
\end{tcolorbox}

\subsection{Motivating Example}
\begin{figure}[!t]
    \centering
    \includegraphics[width=0.98\linewidth]{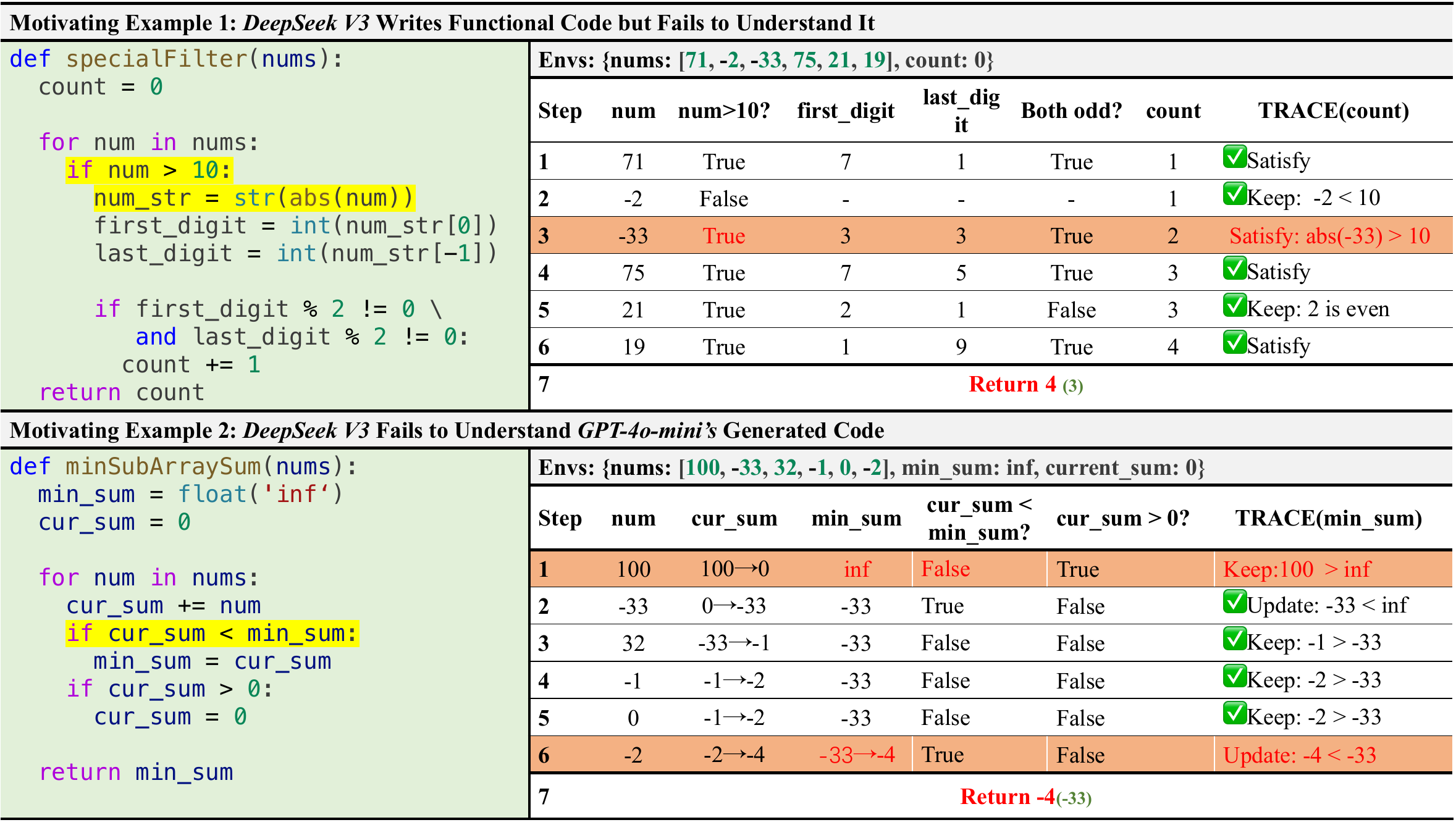}
    \caption{Two motivating examples (\texttt{HumanEval/146} and \texttt{HumanEval/114}) illustrate challenges for code reasoning. The upper-left code counts how many numbers in the input list meet specific digit-based parity conditions, while the bottom-left code finds the minimum sum of any contiguous subarray.
    Although both \textit{DeepSeek-V3} and \textit{GPT-4o-mini} generate logically correct and functional code, the models still make reasoning errors, including logical flaws (upper-right) and mathematical mistakes (bottom-right), highlighted in \textcolor{red}{red}.}
    \label{fig:motivated}
\end{figure}
To concretely illustrate the challenges demonstrated in Sec.\ref{sec:pre-res}, we present two representative cases from HumanEval in Fig.~\ref{fig:motivated} for \emph{intrinsic reasoning limitations} and \emph{self-reasoning bias}, respectively.
In both cases, the code is logically correct by validating against their test cases in a real Python environment, while LLMs fail to reason about their execution paths.

\textbf{Example 1: Intrinsic Reasoning Limitations.} 
The function \texttt{specialFilter} computes how many integers in the input list satisfy a specific digit-based parity condition: (1) The number must be greater than 10. (2) Its first and last digits must both be odd.

Although DeepSeek-V3 correctly generates the code, it fails when reasoning about its own implementation.
Given the input \texttt{nums = [71, -2, -33, 75, 21, 19]}, the model’s reasoning process incorrectly concludes that \texttt{[71, -33, 75, 19]} meets the criteria, yielding a predicted count of 4 instead of the correct 3.
The error occurs at step 3, where the model incorrectly applies the internal \texttt{abs(num)} computation when evaluating \texttt{-33 > 10}, leading it to conclude \texttt{abs(-33) > 10} is true instead of correctly treating the signed value.
However, the comparison \texttt{-33 > 10} should evaluate to \texttt{False} and the subsequent code blocks including \texttt{abs()} should be skipped according to the control flow graph (CFG).

This example exposes limitations of LLMs in deductive code reasoning, as the LLM executor deviates from the corresponding CFGs even in the relatively favorable self-execution scenario, leading to an incorrect branch and producing the wrong result.

\textbf{Example 2: Self-Reasoning Bias.}
The function \texttt{minSubArraySum}, generated by GPT-4o-mini, computes the minimum sum of any contiguous subarray by maintaining a running sum (\texttt{cur\_sum}) and updating the minimum (\texttt{min\_sum}) when a smaller sum is found, resetting \texttt{cur\_sum} to zero whenever it becomes positive.

In this case, GPT-4o-mini generates functionally correct code, but DeepSeek-V3 fails to follow it accurately during execution tracing.
Given the input \texttt{nums = [100, -33, 32, -1, 0, -2]}, the model’s reasoning process deviates from the true execution flow in two places.
At step 1, the comparison \texttt{100 < inf} should evaluate to \texttt{True}, leading to \texttt{min\_sum=100} before resetting \texttt{cur\_sum} to zero.
However, the model incorrectly treats this condition as \texttt{False} and therefore keeps \texttt{min\_sum} as \texttt{inf}, revealing a basic flaw in numerical reasoning.
Later, at step 6, when \texttt{cur\_sum = -4} and \texttt{min\_sum = -33}, the correct evaluation of \texttt{cur\_sum < min\_sum} is \texttt{False}, so \texttt{min\_sum} should remain unchanged.
The model reasoning instead updates \texttt{min\_sum} to \texttt{-4}, introducing a mathematical error.
As a result of these execution mistakes, the final output is \texttt{-4} instead of the correct \texttt{-33}

This reveals that LLMs still suffer from self-reasoning bias, although the LLM reasoner (DeepSeek-V3) is empirically more powerful than the code generator (GPT-4o-mini).

\section{Methodology}
\begin{figure}[!t]
    \centering  \includegraphics[width=0.94\linewidth]{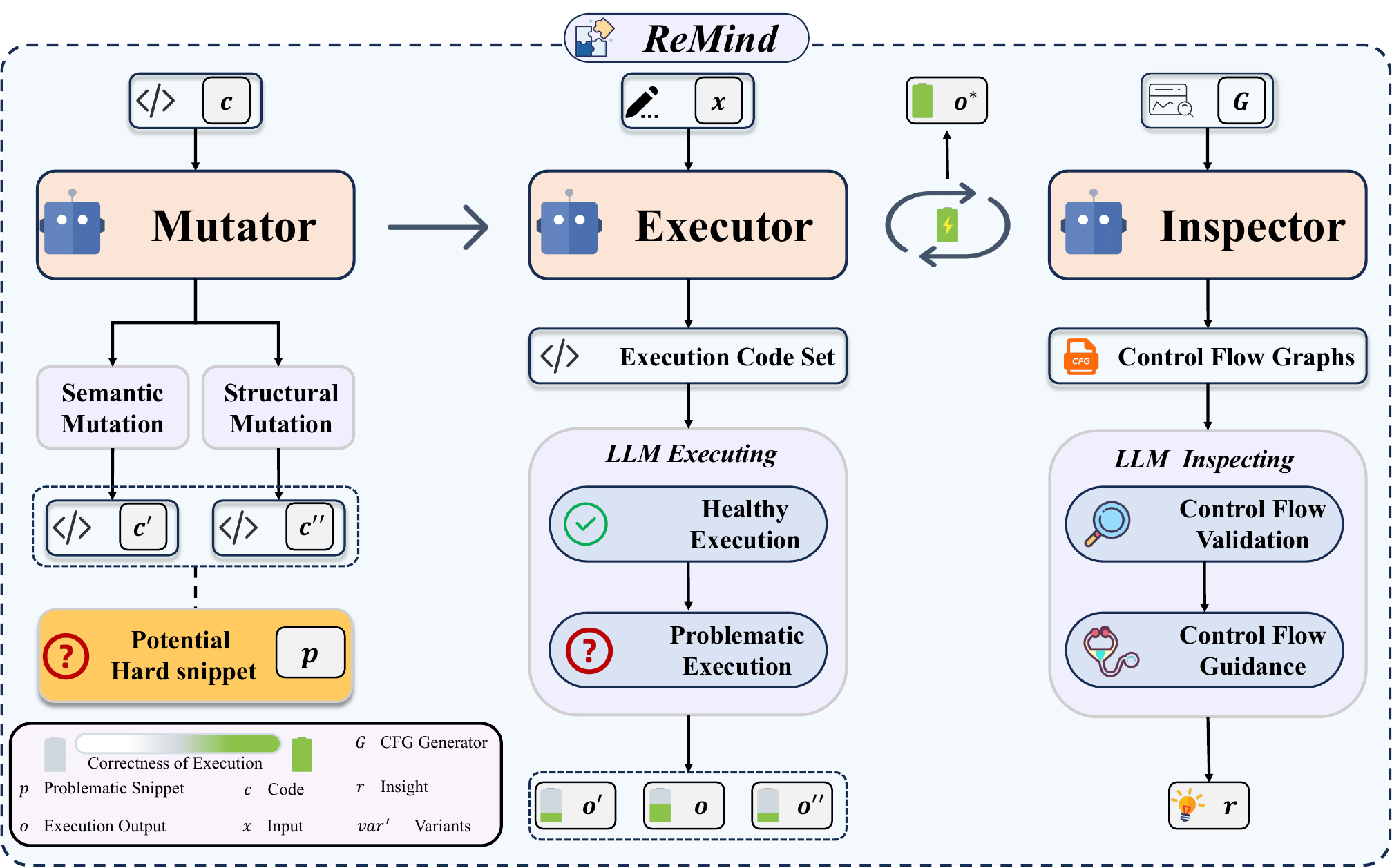}
    \caption{Workflow of the \textbf{\textit{\texttt{ReMind}}} architecture for robust code reasoning.}
    \label{fig:workflow}
\end{figure}

In light of challenges of LLMs in deductive code reasoning, we accordingly propose \texttt{ReMind} in this section, a novel multi-agent framework designed for robust deductive code reasoning by generating diverse code variants, predicting execution paths, and refining problematic trajectories.
As shown in Fig.~\ref{fig:workflow}, \texttt{ReMind} comprises three core components: \texttt{Mutator}, \texttt{Executor}, and \texttt{Inspector}.
In the following sections, we detail the architecture of \texttt{ReMind} and explain how these three agents mitigate intrinsic reasoning gaps, self-reasoning biases, and the zero-shot reasoning dilemma.

\subsection{Architecture}
\subsubsection{Mutator} 
The Mutator plays a central role in alleviating both \emph{intrinsic reasoning gaps} and the \emph{self-reasoning bias} introduced in Section~\ref{sec:pre-res}.
Its primary function is to generate a set of code variants from an original one $c$, thereby enriching the diversity of code representations.
Building on mutation strategies from prior work~\cite{li2024mutation,haroon2025accurately}, it performs two categories of transformations:
\begin{itemize}[leftmargin=*]
    \item \textbf{Semantic Mutation}, which alters code semantics while preserving functional intent (e.g., renaming variables, reordering independent statements).
    \item \textbf{Structural Mutation}, which modifies the structural representation of the program (e.g., rewriting loops into equivalent conditional constructs).
\end{itemize}

Because the Mutator is instantiated from the same LLM as the \texttt{Executor} in \texttt{ReMind}, it naturally produces variants that reflect executor-familiar coding styles.
% Thus, it serves as a bridge for mitigating self-reasoning bias.
Notably, LLM-driven mutations are not always guaranteed to preserve full functional equivalence.
Even for simple snippets involving a few conditionals or loops, SoTA LLMs may fail to fully capture semantic intent as shown in our motivating example in Fig.~\ref{fig:motivated}.
When applied to complex code with mathematical operations or deep branching, mutations can introduce subtle deviations that manifest as inconsistent runtime behaviors~\cite{tip2025llmorpheus,foster2025mutation}.

Rather than discarding these deviations as errors, \texttt{ReMind} treats them as valuable diagnostic signals.
Such deviations highlight code regions that are relatively hard for LLMs to understand and reason about.
These signals are then exposed by the \texttt{Executor} and exploited by the \texttt{Inspector} to identify error-prone fragments and provide refined insights for problematic reasoning traces.
In this way, the \texttt{Mutator} not only produces diverse variants ${c', c''}$ from $c$ to alleviate \emph{self-reasoning bias}, but also exposes implicit reasoning errors that help address the \emph{intrinsic reasoning gap}.

\subsubsection{Executor}
The \texttt{Executor} is responsible for predicting and refining execution trajectories through \emph{behavioral tracing} and \emph{trace refinement}, as illustrated in the middle of Fig.~\ref{fig:workflow}. 
Unlike standard program execution, the \texttt{Executor} in \texttt{ReMind} does not require executability, deterministic interpretation, or sandboxed environments.
Instead, it leverages the reasoning capabilities of LLMs to deduce the execution trajectories of code snippets step-by-step.

Specifically, the \texttt{Executor} begins with \emph{behavioral tracing}, which is initially performed in the absence of feedback from the \texttt{Inspector}. 
In this phase, the \texttt{Executor} directly reasons about both the original and mutated code snippets $\{c, c', c''\}$ for given inputs, producing trajectories $\{o, o', o''\}$ that expose \texttt{Mutator}-introduced deviations and inconsistencies by capturing control-flow decisions, variable updates, and output values.
Next, these initial trajectories are then refined through \emph{trace refinement}, wherein the \texttt{Executor} further incorporates feedback and insights from the \texttt{Inspector}, eventually producing more accurate and reliable trajectories $o^*$.
In general, the execution path predicted by the \texttt{Executor} can be categorized into two types: \textbf{healthy execution} and \textbf{problematic execution}:
\begin{itemize}[leftmargin=*]
    \item \textbf{Healthy execution:} The execution path predicted by \texttt{Executor} aligns with the actual execution of the code.
    
    \item \textbf{Problematic execution:} \texttt{Executor} fails to trace each variable step-by-step and arrive at a wrong answer, often due to the presence of a hard snippet $p$ within $c$. In this paper, problematic executions are further divided into
        \begin{itemize}
            \item \textbf{Explicit errors}: Clearly manifested, as they directly yield incorrect results (see Step 3 of the first motivating example in Fig.~\ref{fig:motivated}).

            \item \textbf{Implicit errors}: More subtle, as their impact is masked by subsequent execution, as shown in Step 1 of the second motivating example in Fig.~\ref{fig:motivated}.
        \end{itemize}
\end{itemize}

\subsubsection{Inspector}
The \texttt{Inspector} constitutes the third component of \texttt{ReMind}, and its primary role is to provide \emph{control-flow–aware validation and guidance} of the trajectories predicted by the \texttt{Executor}.
While the \texttt{Executor} focuses on step-by-step statement tracing, it lacks a principled mechanism to ensure the structural consistency of these predicted traces with the control flows.
The \texttt{Inspector} fills this gap by leveraging \emph{Control Flow Graphs (CFGs)} as a structural condition for validating execution predictions and providing refining insights.

The workflow of the \texttt{Inspector} proceeds in two stages: \emph{control flow validation} and \emph{control flow guidance}.
In the first stage, \emph{control flow validation}, the \texttt{Inspector} calls statistical analysis tools\footnote{We utilized Python’s AST module to parse source code and construct a CFG.} to construct CFGs for both original and mutated snippets.
By horizontally comparing the predicted trajectories ${o, o', o''}$ against the corresponding CFGs, the \texttt{Inspector} identifies inconsistencies between trajectories and control flows, such as infeasible branches, loop misinterpretations, or mismatches between predicted and actual branching conditions, resulting in problematic reasoning.
% These inconsistencies highlight code regions that are hard for LLM reasoning and prone to execution errors.
In the second stage, \emph{control flow guidance}, the \texttt{Inspector} generates repair suggestions that are explicitly aligned with CFG constraints.
These CFG-aligned fixes are returned to the \texttt{Executor} for trace refinement, aiming to produce a more accurate and reliable final execution trace $o^*$.
In this way, it plays a crucial role in alleviating the \emph{intrinsic reasoning gap} by ensuring that predicted execution paths remain both structurally consistent and semantically faithful to the control flow.

Through this dual process, the \texttt{Inspector} leverages the inconsistency introduced by the \texttt{Mutator} and provides feedback to the \texttt{Executor} for trace refinement.
Overall, this closes the loop among mutation, execution, and inspection, enabling \texttt{ReMind} to exhibit outstanding reasoning abilities in both basic and zero-shot settings.

\section{Experiments}
We evaluate the effectiveness and robustness of our proposed approach, \texttt{ReMind}, by comparing it with recent baselines in deductive code reasoning on the tailored benchmark used in the previous empirical study.
In this section, our experimental analysis is structured to answer the following research questions:
\begin{itemize}
    \item \textbf{RQ4:} How effective is \texttt{ReMind} in handling \emph{intrinsic reasoning gap} and \emph{self-reasoning bias} in deductive code reasoning?
    \item \textbf{RQ5:} How does \texttt{ReMind} perform \emph{zero-shot reasoning} on complex, real-world programming benchmarks (e.g., LiveCodeBench)?
    \item \textbf{RQ6:} What is the individual contribution of each \texttt{ReMind} agent (Mutator, Executor, Inspector) to the overall performance?
\end{itemize}

\subsection{Baselines}
We compare our approach against several representative reasoning strategies.
These baselines cover both general-purpose prompting methods (CoT), code-driven prompting methods (CoC), and code-specific reasoning frameworks (RHDA):
\begin{itemize}
    \item  \textbf{CoT (Chain-of-Thought)}~\cite{wei2022chain}, elicits large language models to produce a series of intermediate reasoning steps before arriving at the final answer

    % \item  \textbf{Mutation} extends CoT by introducing semantic mutations via the underlying backbone of LLM Executor before reasoning, to test whether code diversity can alleviate reasoning errors.

    \item  \textbf{CoC (Chain-of-Code)}~\cite{lichain} prompts LLMs to simulate execution by explicitly reasoning over code statements, which enforces more structured tracing than free-form CoT.

    \item  \textbf{RHDA (Reflective Hypothesis Decomposition and Amendment) Pipeline}~\cite{zhaounveiling}, a hierarchical prompting framework, decomposes execution into sub-procedures and aligns reasoning with program structure.
\end{itemize}

\subsection{Main Results}
\begin{table}[!t]
  \centering
\caption{Comparison of accuracy across different executors and prompting methods on the HumanEval benchmark. Darker colors indicate better performance, and the color differences reflect the variation in LLM performance across code origins.}
\begin{tabular}{llccccc|c}
\bottomrule
\multirow{2}{*}{\textbf{Reasoner}} & {\multirow{2}{*}{\textbf{Method}}} & \multicolumn{6}{c}{\textbf{Benchmark: HumanEval}} \\
\cline{3-8}      
&       & 4o-mini & DS-V3 & DS-R1 & o1-Low & o1-High & Avg. \\
\hline
\multirow{4}{*}{\textbf{4o-mini}} & CoT   & \ceco{55.3}  & \ceco{48.7}  & \ceco{51.3}  & \ceco{52.6}  & \ceco{53.9}  & \ceco{52.4} \\
      % & Mutation & -     & \ceco{53.9}  & \ceco{47.4}  & \ceco{53.9}  & \ceco{59.2}  & \ceco{53.6}  \\
      & CoC   & \ceco{61.8}  & \ceco{56.6}  & \ceco{57.9}  & \ceco{57.9}  & \ceco{60.5}  & \ceco{58.9} \\
      & RHDA  & \ceco{63.2}  & \ceco{56.6}  & \ceco{63.2}  & \ceco{64.5}  & \ceco{52.6}  & \ceco{60.0}  \\
      & \texttt{ReMind} & \textbf{\ceco{77.6}} & \textbf{\ceco{72.4}} & \textbf{\ceco{76.3}} & \textbf{\ceco{75.0}} & \textbf{\ceco{77.6}} & \textbf{\ceco{75.8}} \\
\hline
{\multirow{4}{*}{\textbf{DS-V3}}} & CoT   & \ceco{53.9}  & \ceco{85.5}  & \ceco{52.6}  & \ceco{48.7}  & \ceco{55.3}  & \ceco{59.2}   \\
      % & Mutation & \ceco{61.8}  & -     & \ceco{56.6}  & \ceco{53.9}  & \ceco{59.2}  & \ceco{57.9}  \\
      & CoC   & \ceco{59.2}  & \ceco{67.1}  & \ceco{64.5}  & \ceco{65.8}  & \ceco{65.8}  & \ceco{64.5} \\
      & RHDA  & \ceco{65.8}  & \ceco{68.4}  & \ceco{68.4}  & \ceco{71.1}  & \ceco{69.7}  & \ceco{68.7}  \\
      & \texttt{ReMind} & \textbf{\ceco{88.2}} & \textbf{\ceco{96.1}} & \textbf{\ceco{90.8}} & \textbf{\ceco{90.8}} & \textbf{\ceco{92.1}} & \textbf{\ceco{91.6}}  \\
\hline
{\multirow{4}{*}{\textbf{DS-R1}}} & CoT   & \ceco{67.1}  & \ceco{89.5}  & \ceco{92.1}  & \ceco{84.2}  & \ceco{88.2}  & \ceco{84.2}  \\
      % & Mutation & \ceco{78.9}  & \ceco{90.8}  & -     & \ceco{85.5}  & \ceco{90.8}  & \ceco{86.5}   \\
      & CoC   & \ceco{72.4}  & \ceco{75.0}  & \ceco{81.6}  & \ceco{77.6}  & \ceco{75.0}  & \ceco{76.3}   \\
      & RHDA  & \ceco{73.7}  & \ceco{71.1}  & \ceco{75.0}  & \ceco{73.7}  & \ceco{73.7}  & \ceco{73.4}  \\
      & \texttt{ReMind} & \textbf{\ceco{92.1}} & \textbf{\ceco{93.4}} & \textbf{\ceco{98.7}} & \textbf{\ceco{93.4}} & \textbf{\ceco{96.1}} & \textbf{\ceco{94.7}} \\
\hline
\multirow{4}{*}{\textbf{o1-Low}} & CoT   & \ceco{76.3}  & \ceco{81.6}  & \ceco{84.2}  & \ceco{96.1}  & \ceco{92.1}  & \ceco{86.1}   \\
      % & Mutation & \ceco{80.3}  & \ceco{81.6}  & \ceco{88.2}  & -     & \ceco{96.1}  & \ceco{86.5}   \\
      & CoC   & \ceco{75.0}  & \ceco{71.1}  & \ceco{73.7}  & \ceco{80.3}  & \ceco{78.9}  & \ceco{75.8}  \\
      & RHDA  & \ceco{72.4}  & \ceco{72.4}  & \ceco{76.3}  & \ceco{78.9}  & \ceco{77.6}  & \ceco{75.5} \\
      & \texttt{ReMind} & \textbf{\ceco{89.5}} & \textbf{\ceco{93.4}} & \textbf{\ceco{94.7}} & \textbf{\ceco{98.7}} & \textbf{\ceco{98.7}} & \textbf{\ceco{95.0}} \\
\hline
\multirow{4}{*}{\textbf{o1-High}} & CoT   & \ceco{77.6}  & \ceco{78.9}  & \ceco{86.8}  & \ceco{88.2}  & \ceco{97.4}  & \ceco{85.8} \\
      % & Mutation & \ceco{77.6}  & \ceco{85.5}  & \ceco{89.5}  & \ceco{93.4}  & -     & \ceco{86.5}  \\
      & CoC   & \ceco{76.3}  & \ceco{72.4}  & \ceco{77.6}  & \ceco{75.0}  & \ceco{82.9}  & \ceco{76.8}  \\
      & RHDA  & \ceco{71.1}  & \ceco{72.4}  & \ceco{73.7}  & \ceco{75.0}  & \ceco{78.9}  & \ceco{74.2}    \\
      & \texttt{ReMind} & \textbf{\ceco{93.4}} & \textbf{\ceco{92.1}} & \textbf{\ceco{96.1}} & \textbf{\ceco{96.1}} & \textbf{\ceco{100.0}} & \textbf{\ceco{95.5}}  \\
\toprule

\end{tabular}%

\label{tab:exp:main_results}
\end{table}

\begin{table}[!t]
  \centering
  \caption{The standard deviation of accuracy corresponding to \texttt{ReMind} and other baselines in Tab.~\ref{tab:exp:main_results}. Numbers are averaged across different LLMs, where the lower represents the more stable.}
    \begin{tabular}{lcc}
    \toprule
    \multirow{2}{*}{\textbf{Method}} & \multicolumn{2}{c}{\textbf{Benchmark}} \\
          \cmidrule{2-3}
          & \textbf{HumanEval $\downarrow$}    & \textbf{LiveCodeBench $\downarrow$} \\
    \midrule
    CoT   & 8.7   & 3.9  \\
    Mutation & 5.6   & 3.2  \\
    CoC   & 3.3   & 2.9  \\
    RHDA  & \textbf{2.9}   & 2.9  \\
    
    \texttt{ReMind} & \textbf{2.9}   & \textbf{1.3}  \\
    \bottomrule
    \end{tabular}%
  \label{tab:std}%
\end{table}%

\subsubsection{RQ4: How Effective is \texttt{ReMind} in Handling Intrinsic Reasoning Gap and Self-Reasoning Bias in Deductive Code Reasoning?}

From Tab.~\ref{tab:exp:main_results}, we observe that structured reasoning approaches such as CoC and RHDA offer partial improvements. 
On HumanEval, CoC reaches 64.5\% and RHDA achieves 68.7\% with DS-V3, both outperforming plain CoT (59.2\%).
However, their effectiveness diminishes with backbone change to reasoning models like DeepSeek-R1 and o1-Low/High.
Both CoC and RHDA fluctuate around 75\% accuracy, significantly underperforming native CoT prompting (over 84\% accuracy). 
In contrast, our proposed \texttt{ReMind} consistently and significantly outperforms all baselines for any backbones. 
For example, with DeepSeek-V3, \texttt{ReMind} reaches $91.6\%$ on HumanEval, far above CoT ($
59.2\%$), CoC ($64.5\%$), and RHDA ($68.7\%$). 
% These results highlight that while CoT, CoC, and RHDA provide incremental gains, they remain fragile under different settings, whereas \texttt{ReMind} delivers robust and superior accuracy.
More importantly, \texttt{ReMind} eliminates \emph{self-reasoning bias}. 
While baselines often show large accuracy gaps between different executor backbones.
For example, CoT with DeepSeek-V3 achieves $85.5\%$ accuracy while only achieving $48.7\%$ accuracy when reasoning over code generated by o1-Low.
While \texttt{ReMind} consistently sustains high robustness desregarding code sources, as shown in Tab.~\ref{tab:std} and color differences in Tab.~\ref{tab:exp:main_results} and Tab.~\ref{tab:exp:main_results_lcb}. 
% This consistency demonstrates that our design does not merely amplify reasoning accuracy but fundamentally alleviates the self-reasoning bias of LLMs, yielding stable and outstanding reasoning performance.

Experimental results indicate that although CoC and RHDA offer improvements over CoT in certain settings, they still rely on ``one-shot correctness'' and fall short in refinement.
Once an error enters the reasoning trace, it is rarely detected or corrected.
CoC enforces more structured tracing, but it is fragile across different code sources and fails to refine reasoning mistakes because it operates as an end-to-end prompting method similar to CoT.
RHDA incorporates reflection and decomposition, which improves robustness to different code styles.
However, its reflection mechanism depends solely on the internal ability of LLMs without external guidance, making it hard to recognize and refine problematic reasoning.
Specifically, in Fig.~\ref{fig:casestudy}, the analysis of RHDA has pointed out the execution flow, while its reasoning process still goes wrong. 
In contrast, ReMind’s strength lies in its ability to identify and repair reasoning errors.
The Mutator breaks the \emph{self-reasoning bias}, the Inspector explores diverse execution trajectories, and the Inspector enforces CFG alignment to pull reasoning back onto valid paths. 
This closed-loop design explains why \texttt{ReMind} not only achieves higher absolute accuracy but also substantially reduces performance variance across code origins.

\begin{tcolorbox}[breakable,width=\linewidth-2pt,boxrule=0pt,top=2pt, bottom=2pt, left=2pt,right=2pt, colback=gray!20,colframe=gray!20]
\textbf{Answer to RQ4:} \texttt{ReMind} effectively mitigates \emph{intrinsic reasoning gap} and \emph{self-reasoning bias}, achieving outstanding accuracy and robustness compared with all baselines.
\end{tcolorbox}

\begin{table}[!t]
  \centering
\caption{Comparison of accuracy across different executors and prompting methods on the LiveCodeBench benchmark.}

\begin{tabular}{llccccc|c}
\bottomrule
\multirow{2}{*}{\textbf{Reasoner}} & {\multirow{2}{*}{\textbf{Method}}} &  \multicolumn{6}{c}{\textbf{Benchmark: LiveCodeBench}} \\
\cline{3-8}      
&         & 4o-mini & DS-V3 & DS-R1 & o1-Low & o1-High & Avg. \\
\hline
\multirow{4}{*}{\textbf{4o-mini}} & CoT    & \ceco{42.1}  & \ceco{34.8}  & \ceco{40.2}  & \ceco{36.6}  & \ceco{40.9}  & \ceco{38.9}  \\
      % & Mutation   & -     & \ceco{35.4}  & \ceco{40.9}  & \ceco{41.5}  & \ceco{42.7}  & \ceco{40.1}  \\
      & CoC   & \ceco{44.5}  & \ceco{39.0}  & \ceco{39.6}  & \ceco{41.5}  & \ceco{39.0}  & \ceco{40.7}  \\
      & RHDA  & \ceco{39.0}  & \ceco{35.4}  & \ceco{37.8}  & \ceco{38.4}  & \ceco{37.2}  & \ceco{37.6}  \\
      & \texttt{ReMind} & \textbf{\ceco{47.0}} & \textbf{\ceco{48.8}} & \textbf{\ceco{50.6}} & \textbf{\ceco{49.4}} & \textbf{\ceco{50.0}} & \textbf{\ceco{49.1}} \\
\hline
{\multirow{4}{*}{\textbf{DS-V3}}} & CoT   & \ceco{39.0}  & \ceco{48.8}  & \ceco{41.5}  & \ceco{36.9}  & \ceco{42.7}  & \ceco{41.8}  \\
      % & Mutation & \ceco{61.8}   & -     & \ceco{43.3}  & \ceco{47.6}  & \ceco{50.6}  & \ceco{47.3}  \\
      & CoC   & \ceco{49.4}  & \ceco{53.7}  & \ceco{51.2}  & \ceco{51.2}  & \ceco{51.8}  & \ceco{51.5}  \\
      & RHDA   & \ceco{48.8}  & \ceco{57.9}  & \ceco{49.4}  & \ceco{48.8}  & \ceco{49.4}  & \ceco{50.9}  \\
      & \texttt{ReMind} & \textbf{\ceco{67.1}} & \textbf{\ceco{70.1}} & \textbf{\ceco{68.3}} & \textbf{\ceco{68.9}} & \textbf{\ceco{69.5}} & \textbf{\ceco{68.8}} \\
\hline
{\multirow{4}{*}{\textbf{DS-R1}}} & CoT    & \ceco{50.0}  & \ceco{54.3}  & \ceco{57.3}  & \ceco{54.3}  & \ceco{54.9}  & \ceco{54.1}  \\
      % & Mutation   & \ceco{53.7}  & \ceco{55.5}  & -     & \ceco{54.9}  & \ceco{56.7}  & \ceco{55.2}  \\
      & CoC   & \ceco{51.2}  & \ceco{52.4}  & \ceco{57.9}  & \ceco{54.9}  & \ceco{56.1}  & \ceco{54.5}  \\
      & RHDA   & \ceco{52.4}  & \ceco{51.8}  & \ceco{57.3}  & \ceco{53.0}  & \ceco{53.7}  & \ceco{53.7}  \\
      & \texttt{ReMind}  & \textbf{\ceco{76.2}} & \textbf{\ceco{76.8}} & \textbf{\ceco{78.7}} & \textbf{\ceco{78.0}} & \textbf{\ceco{78.7}} & \textbf{\ceco{77.7}} \\
\hline
\multirow{4}{*}{\textbf{o1-Low}} & CoT   & \ceco{45.7}  & \ceco{56.7}  & \ceco{56.7}  & \ceco{61.6}  & \ceco{57.9}  & \ceco{55.7}  \\
      % & Mutation  & \ceco{47.0}  & \ceco{59.1}  & \ceco{57.9}  & -     & \ceco{59.8}  & \ceco{55.9}  \\
      & CoC   & \ceco{48.8}  & \ceco{53.7}  & \ceco{55.5}  & \ceco{59.1}  & \ceco{56.7}  & \ceco{54.8}  \\
      & RHDA   & \ceco{50.6}  & \ceco{53.7}  & \ceco{54.3}  & \ceco{56.7}  & \ceco{61.0}  & \ceco{55.2}  \\
      & \texttt{ReMind}  & \textbf{\ceco{72.6}} & \textbf{\ceco{73.2}} & \textbf{\ceco{73.8}} & \textbf{\ceco{74.4}} & \textbf{\ceco{75.0}} & \textbf{\ceco{73.8}} \\
\hline
\multirow{4}{*}{\textbf{o1-High}} & CoT    & \ceco{54.9}  & \ceco{57.3}  & \ceco{59.1}  & \ceco{60.1}  & \ceco{64.0}  & \ceco{59.1}  \\
      % & Mutation   & \ceco{56.1}  & \ceco{59.8}  & \ceco{60.4}  & \ceco{62.8}  & -     & \ceco{59.8}  \\
      & CoC    & \ceco{53.0}  & \ceco{55.5}  & \ceco{57.3}  & \ceco{58.5}  & \ceco{63.4}  & \ceco{57.6}  \\
      & RHDA   & \ceco{57.3}  & \ceco{59.8}  & \ceco{60.4}  & \ceco{62.8}  & \ceco{65.2}  & \ceco{61.1}  \\
      & \texttt{ReMind}  & \textbf{\ceco{76.2}} & \textbf{\ceco{77.4}} & \textbf{\ceco{78.0}} & \textbf{\ceco{79.3}} & \textbf{\ceco{80.8}} & \textbf{\ceco{78.4}} \\
\toprule

\end{tabular}%

\label{tab:exp:main_results_lcb}
\end{table}

\subsubsection{RQ5: How Does \texttt{ReMind} Perform Zero-shot Reasoning on Complex, Real-world Programming Benchmarks?}

LiveCodeBench represents a substantially harder benchmark compared to HumanEval, involving larger code contexts, real-world library usage, and complex execution dependencies. 
Thus, all the methods suffer severe generalization drops when moving to this benchmark due to the knowledge cutoff LLMs. 
For instance, with DeepSeek-R1, CoT drops from $84.2\%$ (HumanEval) to $54.1\%$ (LiveCodeBench), CoC decreases from $76.3\%$ to $54.5\%$, and RHDA decreases from $73.4\%$ to $53.7\%$ as shown in Tab.~\ref{tab:exp:main_results_lcb}.    

In contrast, \texttt{ReMind} still maintains strong performance on LiveCodeBench, consistently outperforming baselines by large margins. 
For example, with o1-High, \texttt{ReMind} achieves $95.5\%$ on HumanEval and still retains $78.4\%$ accuracy on LiveCodeBench, significantly higher than CoT ($59.1\%$), CoC (57.6\%), and RHDA ($61.1\%$). 
% This strongly highlights the scalability of our method, which \texttt{ReMind} not only excels in standard benchmarks but also sustains reasoning accuracy under complex, real-world programming conditions.
These results demonstrate the practicality, generalization strength, and scalability of \texttt{ReMind}, confirming its capability for reliable zero-shot reasoning rather than being tailored to specific benchmarks.

\begin{tcolorbox}[breakable,width=\linewidth-2pt,boxrule=0pt,top=2pt, bottom=2pt, left=2pt,right=2pt, colback=gray!20,colframe=gray!20]
\textbf{Answer to RQ5:} On LiveCodeBench, \texttt{ReMind} sustains strong zero-shot performance while other methods degrade sharply, confirming its scalability to complex, real-world programming tasks.
\end{tcolorbox}

\subsubsection{RQ5: What is the individual contribution of each \texttt{ReMind} agent to the overall performance?}

\begin{table}[!t]
  \centering
  \caption{Ablation study of \texttt{ReMind} over Mutator (M) and Inspector (I) in Self-Execution settings.}
  \resizebox{\textwidth}{!}{
% Table generated by Excel2LaTeX from sheet 'Sheet2'
\begin{tabular}{lccccc|ccccc}
\bottomrule
\multirow{2}{*}{\textbf{Method}} & \multicolumn{5}{c|}{\textbf{Benchmark: HumanEval}} & \multicolumn{5}{c}{\textbf{Benchmark: LiveCodeBench}} \\
\cline{2-11}      & 4o-mini & DS-V3 & DS-R1 & o1-Low & o1-High & 4o-mini & DS-V3 & DS-R1 & o1-Low & o1-High \\
\hline
\texttt{ReMind} & 77.6  & 96.1  & 98.7  & 98.7  & 100.0  & 47.0  & 70.1  & 78.7  & 74.4  & 80.8  \\
\quad w/o M & 68.4  & 90.1  & 96.1  & 97.4  & 98.7  & 44.7  & 58.0  & 67.1  & 68.9  & 73.2  \\
\quad w/o I & 61.8  & 87.5  & 94.7  & 96.7  & 98.7  & 43.4  & 54.9  & 64.6  & 65.8  & 69.5  \\
\quad w/o I + M & 55.3  & 85.5  & 92.1  & 96.1  & 97.4  & 42.1  & 48.8  & 57.3  & 61.6  & 64.0  \\
\toprule
\end{tabular}%

    }
  \label{tab:ablation}%
\end{table}%

To better understand the contribution of each component in \texttt{ReMind}, we conduct ablation studies by selectively removing the Mutator (M) and Inspector (I) agents in self-execution settings.
Specifically, when removing the Mutator (M), the Executor reasons only over the original code, without semantic or structural mutations.
In this case, the Inspector still operates, but its refinements are based solely on the CFG of the unaltered snippets and its execution trajectory.
In contrast, removing the Inspector (I) means that Mutator is still active, so both the original and mutated code are executed, but the final answer is determined by majority voting without any CFG-driven inspection or correction.
Removing both Mutator and Inspector represents that \texttt{ReMind} degrades to native CoT prompting.

The results in Tab.~\ref{tab:ablation} yield three main findings:
\begin{itemize}
    \item \textbf{Role of Mutator (M).} Removing the Mutator leads to consistent drops in accuracy for both HumanEval and LiveCodeBench. 
    This confirms that code mutations are essential in diversifying execution traces and exposing mistakes during reasoning.
    The impact of the Mutator is particularly pronounced on more challenging benchmarks.
    For instance, with DeepSeek-V3 on LiveCodeBench from $70.1\%$ to $58.0\%$, and DeepSeek-R1 drops from 78.7\% to 67.1\%.
    This demonstrates that, in addition to alleviating self-reasoning bias, the Mutator plays a vital role in identifying potentially difficult code segments, especially in complex coding tasks.

    \item \textbf{Role of Inspector (I).} Excluding the Inspector yields even larger degradations. 
    For example, with DeepSeek-V3, accuracy falls from $70.1\%$ to $54.9\%$ in LiveCodeBench. 
    This underscores the Inspector's importance in identifying and correcting flawed reasoning paths through CFg-driven analysis. 
    Notably, the performance decline when removing the Inspector is consistently greater than when removing the Mutator. 
    This indicates that while the Mutator effectively surfaces error-prone code segments, the Executor cannot fully leverage this diversity, even using majority voting to determine final results.
    Hence, the Inspector is crucial for refining flawed reasoning paths through CFG-driven analysis.
    \item \textbf{Collaboration of Agents.} When both modules are removed simultaneously, \texttt{ReMind} degrades to the native CoT prompting, leading to a more severe performance drop than when either the Mutator or Inspector is removed individually.
    For instance, with o1-High on LiveCodeBench, accuracy drops from $80.8\%$ to $64.0\%$.
    This highlights the complementary roles of the two agents, where the Mutator broadens the search space, and the Inspector ensures correctness and stability, jointly yielding the robustness of \texttt{ReMind}.
\end{itemize}

Overall, these ablation studies validate the design of \texttt{ReMind}. 
While each component offers individual benefits, the complete \texttt{ReMind} pipeline is essential to achieve outstanding performance and robustness across both basic and zero-shot settings.

\begin{tcolorbox}[breakable,width=\linewidth-2pt,boxrule=0pt,top=2pt, bottom=2pt, left=2pt,right=2pt, colback=gray!20,colframe=gray!20]
\textbf{Answer to RQ6:} Both Mutator and Inspector are indispensable. The former diversifies execution traces and exposes error-prone code segments, while the latter performs CFG-driven corrections. Their synergy is key for \texttt{ReMind} to achieve both correctness and robustness.
\end{tcolorbox}

% \subsubsection{Summary of Observations.}

% Overall, the experiments demonstrate that \texttt{ReMind} consistently outperforms existing baselines across both HumanEval and LiveCodeBench, achieving higher accuracy and significantly alleviating \emph{intrinsic reasoning gap} and \emph{self-reasoning bias}.
% Unlike CoT, Mutation, CoC, and RHDA, whose performance varies significantly with different executors and benchmarks, \texttt{ReMind} maintains stable and strong results across all settings (\emph{RQ4}). 
% On the more challenging LiveCodeBench benchmark, it further shows strong generalization ability.
% \texttt{ReMind} sustains high accuracy where other methods severe significant performance degradation, which highlights its practicality in complex, real-world programming scenarios (\emph{RQ5}).
% Finally, the ablation studies confirm that both the Mutator and Inspector agents are indispensable in \texttt{ReMind}.
% The Mutator enhances coverage through diverse code variants, while the Inspector leverages CFG-driven analysis to correct reasoning errors.
% Removing either component leads to performance drops, but removing both reverts the framework to plain CoT and results in the most significant degradation, highlighting the synergistic effect of their collaboration. (\emph{RQ6}).

% Collectively, these findings confirm that \texttt{ReMind} is a robust, generalizable, and outstanding framework for code reasoning.

\section{Discussion}
\subsection{Case Study}
\begin{figure}[!t]
    \centering
    \includegraphics[width=0.98\linewidth]{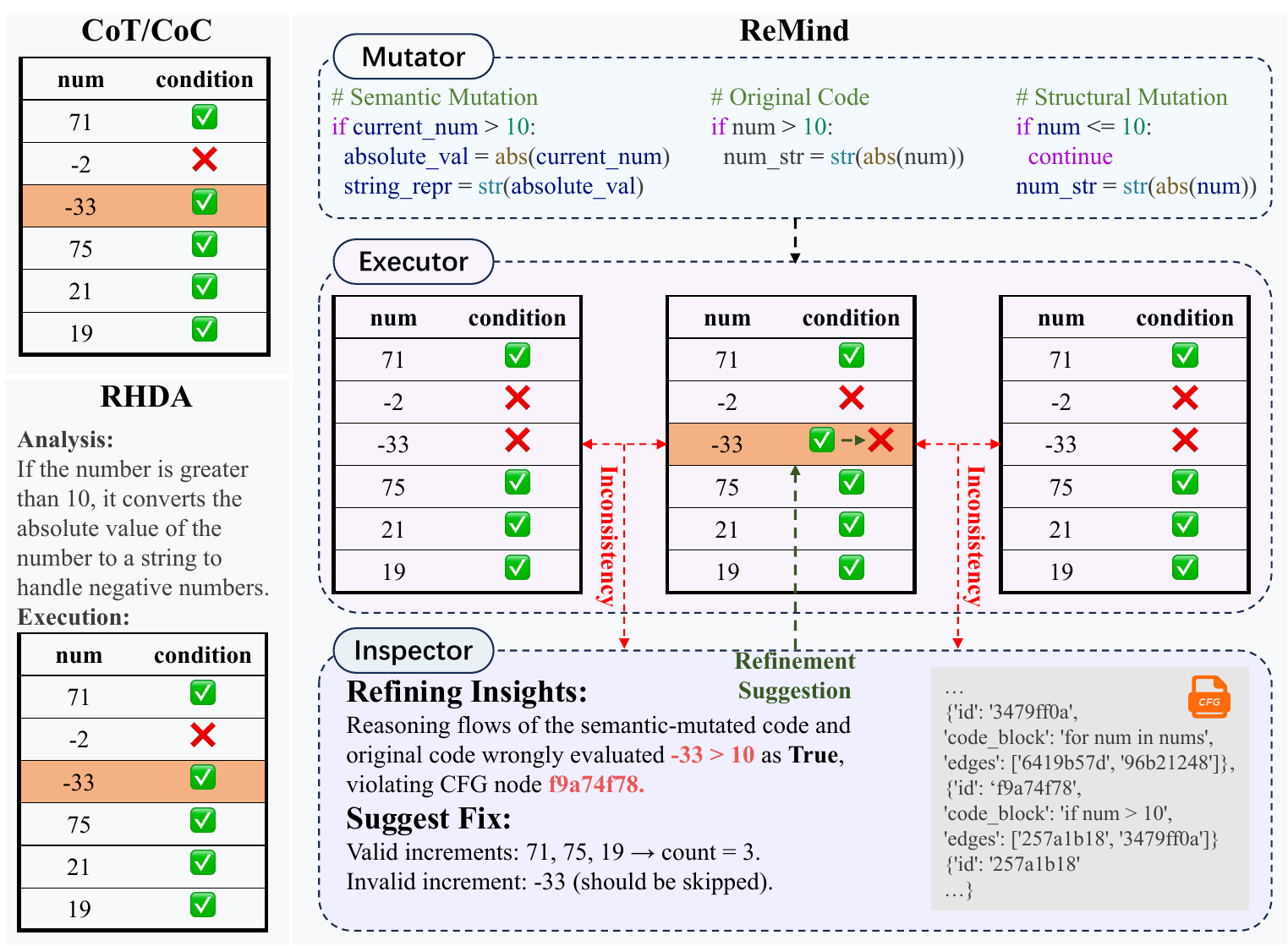}
    \caption{A case study of \texttt{HumanEval/146} demonstrating how \texttt{ReMind} identifies and corrects a reasoning error through the collaboration of Mutator, Executor, and Inspector. Original code is generated by DeepSeek-V3, while o1-high serves as the underlying LLM of \texttt{ReMind}.}
    \label{fig:casestudy}
\end{figure}
As shown in Fig.~\ref{fig:casestudy}, this case study demonstrates how our \texttt{ReMind} framework identifies and corrects a reasoning error in a generated program by analyzing a critical control flow condition.
The full code originates from the \texttt{specialFilter} function in the motivating example 1 introduced earlier in Fig.~\ref{fig:motivated}, which involves processing a list of integers to count how many satisfy a given condition.
For clarity, we focus here only on the problematic region of the code where the error occurs.
According to the intended logic, only values strictly greater than 10 should be processed and counted.
However, during execution tracing of the original code, the LLM incorrectly treats \texttt{num = -33} as satisfying the condition \texttt{num > 10}, similar to other baselines, leading to an invalid increment of the counter.

To identify and resolve this issue, the Mutator first generates semantically and structurally valid variants of the conditional block.
These variants include logically equivalent rewrites of the condition (e.g., replacing \texttt{num > 10} with not (\texttt{num <= 10})), renaming variables (e.g., renaming \texttt{num} to \texttt{current\_num}), or adjusting control flow structure while preserving intended behavior (e.g., \texttt{continue}).
These mutations expand the diversity of code styles and help expose hidden reasoning flaws that may not be apparent in the original formulation.
Each mutated version is then passed to the \texttt{Executor}, which reasons about the condition of the original code and its variants over the input list \texttt{nums = [71, -2, -33, 75, 21, 19]}.
The \texttt{Inspector} subsequently validates the execution traces against their CFGs.
By comparing three regions involving \texttt{num = -33} in each trace, the \texttt{Inspector} flags the transition of the original code at node \texttt{f9a74f78} as invalid.
The condition \texttt{num > 10} is evaluated as true despite num being negative, which contradicts basic arithmetic semantics.
This inconsistency is detected precisely, and the \texttt{Inspector} then provides a clear refinement suggestion back to the \texttt{Executor} that -33 should be skipped as only numbers strictly greater than 10 should be counted.
Based on this, the trace is refined, and the correct count is updated to 3.

In summary, the Mutator generates diverse code variants to mitigate self-execution bias and surface potential reasoning errors.
The Executor then traces each variant step-by-step, producing detailed reasoning traces.
Finally, the Inspector verifies these traces against the corresponding CFGs, identifying inconsistencies and providing actionable feedback to refine the original reasoning.
This closed-loop process enables robust error detection and refinement through the collaborative integration of mutation, execution, and inspection, improving both the reliability and accuracy of deductive code reasoning.

Fig.~\ref{fig:casestudy} also illustrates how CoT, CoC, and RHDA reason about the condition.
End-to-end prompting methods like CoT and CoC wrongly evaluate \texttt{-33 > 10} as true without any further refinements.
Among them, RHDA correctly recognizes that the reasoning should first check whether the number is larger than 10 and then convert it to its absolute value.
However, the reasoning process remains imperfectly refined, indicating that the reasoning process fails to leverage analysis to arrive at the correct answer.

\subsection{Test-time Scaling Costs}
% \begin{table}[!t]
%   \centering
%   \caption{Avg. API calls using o1-High across HumanEval and LiveCodeBench.}
%     \begin{tabular}{lcc}
%     \toprule
%     \multirow{2}{*}{\textbf{Method}} & \multicolumn{2}{c}{\textbf{Avg. API Calls}} \\
% \cmidrule{2-3}          & HumanEval & LiveCodeBench \\
%     \midrule
%     CoT   & 1.0   & 1.0  \\
%     Mutation & 2.0   & 2.0  \\
%     CoC   & 1.0   & 1.0  \\
%     RHDA  & 5.3   & 5.5  \\
%     \texttt{ReMind} & 5.0   & 5.2  \\
%     \bottomrule
%     \end{tabular}%
%   \label{tab:costs}%
% \end{table}%

% We report the average number of API calls across benchmarks.
According to experimental statistics, \texttt{ReMind} requires 5.0 and 5.2 API calls for HumanEval and LiveCodeBench on average, which is higher than end-to-end prompting methods such as CoT and CoC but lower than RHDA, which relies on iterative refinement and requires 5.3 and 5.5 API calls, respectively.
This increased cost of \texttt{ReMind} stems primarily from generating multiple mutated code variants and performing independent reasoning over each.
Furthermore, inconsistencies in reasoning traces trigger the Inspector to analyze discrepancies and provide refinement suggestions for the Executor to revise its prior reasoning.

We view this as a deliberate trade-off under the paradigm of test-time scaling~\cite{jaech2024openai,muennighoff2025s1,zeng2025revisiting,gao2025aim,gao2025uniicl,gao2025interleaved}.
The additional API calls represent allocated compute resources dedicated to mutation, execution, and inspection.
By leveraging more inference-time computation, \texttt{ReMind} enhances the reliability of deductive code reasoning.
Thus, the incurred cost constitutes a structured investment in reasoning quality, aligning with the principle that performance can be improved by scaling computation during inference.

\subsection{Threats to Validity}
While \texttt{ReMind} demonstrates significant improvements in the \emph{intrinsic reasoning limitations} of LLMs and mitigates \emph{self-execution bias} across diverse LLMs in deductive code reasoning, several limitations remain and point to important directions for future work.

First, our empirical evidence reveals that while \texttt{ReMind} enhances reasoning robustness, its performance is ultimately constrained by the base model’s ability to simulate program execution and comprehend complex control flows.
Weaker models may struggle to benefit fully from the mutation and inspection processes, especially in handling deeply nested logic or recursion.
Second, the current implementation of \texttt{ReMind} is primarily designed for sequential, deterministic programs.
It may face challenges when applied to concurrent, asynchronous, or side-effect-heavy code (e.g., I/O operations, network calls), where the runtime behavior is harder to simulate without actual execution environments.
Third, the computational overhead introduced by the multi-agent workflow, including code mutation, execution tracing, and CFG inspection, may be heavier than end-to-end prompting methods.

We believe that addressing these limitations represents a promising path toward even more robust and scalable code reasoning frameworks.

\section{Related Work}
\subsection{Code Reasoning with LLMs}
Large language models (LLMs) have made significant progress in code-related tasks~\cite{liu2024deepseek,guo2025deepseek,jaech2024openai,dong2025survey}, yet they often fall short in robust code understanding and reasoning~\cite{gucruxeval,lichain,zhaounveiling,jainlivecodebench,tian2025codehalu}.
Effective code reasoning demands not only semantic comprehension but also awareness of syntactic and structural context~\cite{tian2025codehalu}.
In light of this challenge, recent approaches leverage execution feedback and iterative refinement that models generate code, observe runtime outcomes, and accordingly revise their solutions~\cite{xue2024selfpico,zhaounveiling}.
Other strategies employ structured prompting or role-based constraints to improve coherence and task alignment~\cite{hong2024metagpt}.
Nonetheless, performance is frequently limited by insufficient contextual awareness, particularly in complex or long-horizon coding tasks~\cite{bi2024iterative,yang2025code}.

\subsection{LLM-Driven Multi-Agent Frameworks}
To tackle the complexity of real-world software development, LLM-driven multi-agent frameworks have emerged as a powerful paradigm~\cite{guo2025repoaudit,mo2025interactive,cihon2025measuring} in both industry and academia~\cite{wang2023survey,zhang2023igniting}.
These systems assign specialized roles, such as planner, coder, and manager, to distinct agents, enabling collaborative problem solving through structured interaction~\cite{chen2025investigating,yue2025masrouter,hong2024metagpt}.
Inspired by human organizational workflows, they utilize standardized operating procedures to coordinate agent behavior and maintain output consistency~\cite{hong2024metagpt}.
By decomposing tasks and distributing cognitive load, multi-agent frameworks offer enhanced modularity, interpretability, and robustness in end-to-end software engineering scenarios.

\section{Conclusion}
In this paper, we investigated the limitations of large language models in deductive code reasoning, the ability to predict program runtime behavior through pure reasoning without actual execution.
Through an empirical study, we identified two critical challenges: \emph{(1) LLMs often fail to accurately reason about the execution of even functionally correct code they themselves generated,} and (2) \emph{they exhibit a strong self-execution bias, performing significantly better on code generated by their own model than on logically equivalent code from other sources.}
To address these issues, we proposed \texttt{ReMind}, a novel multi-agent framework that integrates code mutation, behavioral tracing, and control-flow inspection to enhance reasoning accuracy and robustness.
\texttt{ReMind} enables systematic exploration of code variants, simulates execution traces, and validates reasoning paths against control flow graphs to detect and correct flaws.
Extensive experiments on basic and complex benchmarks demonstrated that \texttt{ReMind} significantly improves code reasoning accuracy across diverse LLMs, reduces self-execution bias, and enhances zero-shot generalization.
\texttt{ReMind} advances the reliability of using LLMs in critical software engineering applications such as automated debugging, program verification, and AI-assisted development. We believe \texttt{ReMind} provides a promising direction toward building more trustworthy and robust code-reasoning systems.

\section{Data Availability}
The code and data are available at \url{https://anonymous.4open.science/r/remind-71F0/}
\bibliographystyle{ACM-Reference-Format}
\bibliography{sample-base}

\appendix
\end{document}